**Diamond nanophotonic circuits functionalized by dip-pen nanolithography**

By *Patrik Rath[1,\*], Michael Hirtz[1,§,\*], Georgia Lewes-Malandrakis[2], Dietmar Brink[2], Christoph Nebel[2], and Wolfram H. P. Pernice[1,&]*

[1]  Patrik Rath, Michael Hirtz, Wolfram H. P. Pernice
     Institute of Nanotechnology (INT) & Karlsruhe Nano Micro Facility (KNMF), Karlsruhe Institute of Technology, Hermann-von-Helmholtz-Platz 1, 76344 Eggenstein-Leopoldshafen (Germany)
[2]  Georgia Lewes-Malandrakis, Dietmar Brink, Christoph Nebel
     Fraunhofer Institute for Applied Solid State Physics, Tullastr. 72, 79108 Freiburg (Germany)
[\*] These authors contributed equally.
[§]  E-mail: michael.hirtz@kit.edu
[&]  E-mail: wolfram.pernice@kit.edu

Integrated nanophotonic circuits made from wide-bandgap semiconductors offer exciting prospects for advanced sensing applications and broadband optical data processing. Among the available substrates, diamond is particularly appealing due to its long-term stability, biocompatibility and chemical inertness. Because of strong field confinement in diamond waveguides, near-field effects can be efficiently harnessed to realize building blocks for biofunctional circuits in a scalable fashion. Here, we report on the parallel and site-specific functionalization of diamond nanophotonic devices as a promising route towards waferscale bio-photonic systems. We show that arbitrary geometric features can be surface modified using dip-pen nanolithography with a minimum linewidth of 100 nm. We simultaneously functionalize several microring and microdisc resonators with different dies and high precision, allowing us to route fluorescent emission with photonic waveguides to arbitrary locations on chip. Our approach holds promise for hybrid optical systems and nanoscale bioactive devices for robust biomedical and environmental high-throughput sensing applications.

Photonic components made from diamond have emerged as a promising platform for applications in quantum optics [1-4], non-linear optics [5,6] and optomechanics [7,8]. Because of its remarkable material properties such as broadband optical transparency, high mechanical stability and hardness, high thermal conductivity, and good chemical stability, diamond is used for a wealth of applications in research and industrial environments. In particular the combination of appealing optical properties and biocompatibility make diamond an attractive platform for biophotonic applications [9-11]. Nowadays high quality diamond thin films are available on large-scale substrates by relying chemical vapor deposition (CVD) which allows for using waferscale processing techniques to realize functional devices. In this way homogeneous large scale diamond thin films can be achieved (see Supplementary Fig. S2). Using nanofabrication routines developed originally for the electronics industry CVD diamond thin films can be structured into subwavelength photonic devices, which allow for propagating light over centimeter distances, the realization of high quality optical resonators and on-chip interferometers [8,12]. This way the powerful toolbox available to integrated optics can be efficiently utilized to realize a sensitive readout platform for biological signals.

In order to employ integrated photonic devices for biosensing suitable links to desired analytes have to be established. Several techniques are commonly employed to functionalize pre-structured photonic substrates, including spincoating with suitable coatings introduced in the liquid phase [13,14] and exposure to reactive agents in the gas phase [15]. While such techniques are suitable to functionalize many photonic components uniformly, it is ultimately necessary to only target specific locations on chip. Deposition of functional and bio-functional patterns has been demonstrated using various lithography techniques [16-18]. Among the available options, dip-pen nanolithography (DPN) has recently been established as a promising technique for reaching spatial resolution compatible with the spatial length scales which nanophotonic devices typically occupy. DPN [19] was first used to deposit thiols on gold surfaces but was quickly adapted for a wide range of ink materials, including biological inks where the mild process parameters are of special value for keeping these often delicate compounds intact [20]. DPN employs a direct-writing method to place "inks" such as small organic molecules [21,22], polymers [23,24], biomolecules [25,26], or nanoparticles [27,28] onto a solid substrate using an atomic force microscopy (AFM) tip as a "pen". The use of phospholipids and lipid mixtures in DPN, termed Lipid-DPN (L-DPN), enables site specific and multiplexed (i.e. different inks written at the same time and in close vicinity on the surface) functionalization of surfaces with biomimetical membrane stacks used e.g. in immunological assays [29-32]. This technique does not rely on a specific surface chemistry and can therefore be applied to various substrates, e.g. silicon oxide, polymers [29], graphene [33] and recently we demonstrated precise functionalization of three dimensional goblet sensor structures by L-DPN [34]. The challenge is to combine the parallel functionalization of devices with different inks with the functionalization of pre-structured materials with good alignment, opposed to bare flat substrates. Furthermore the DPN functionalization of photonic structures from a wide-gap semiconductor is a concept with interesting potential applications, which has not been shown yet.

Here, we employ L-DPN for site-specific functionalization of waferscale diamond integrated optical circuits using lipid mixtures, with high precision and line resolution down to 100nm. Using waferscale, transparent diamond thin films deposited by CVD followed by chemo-mechanical polishing we realize nanophotonic circuitry with low-surface roughness, suitable for later surface modification. We combine planar nanofabrication techniques with DPN functionalization in order to create biocompatible integrated optical circuits that are locally modified for biological functionality. With L-DPN we directly write onto fabricated straight optical waveguides and curved ring resonators. Using a multi-pen strategy, several devices are functionalized in parallel, each with different modifiers. With fluorescence microscopy and atomic force microscopy we verify the site-specific surface modification and demonstrate guiding and coupling of fluorescent light in nanophotonic waveguides. Additionally, we demonstrate the delivery of ink solutions by microchannel cantilever spotting [35, 36], exploiting a self-alignment process for functionalization of ring resonators as an alternative route for solvent based delivery of active material. Our approach holds promise for high throughput and high accuracy fabrication of biophotonic devices, which will enable applications in bio-sensing and bio-interfacing by the multiplexed introduction of active elements in form of functional ink mixtures.

While traditional semiconductor materials are frequently employed for the fabrication of nanophotonic components diamond devices promise several advantages over existing material templates. The large electronic bandgap enables optical transparency above 220 nm all the way into the long infrared spectral region and strongly suppresses two-photon absorption commonly encountered in silicon substrates. Because diamond is carbon based, covalent binding can be utilized to link its surface to suitable biomarkers [37]. In addition, because diamond has been shown to provide excellent biocompatibility, devices made from diamond are a promising platform for biomedical applications [38]. Here we employ CVD diamond thin films, which can be used to tightly confine light at the nanoscale because of a large refractive index of 2.4 (see Supplementary Fig. S3 for the waveguide geometry and guided modes). About 900 nm of polycrystalline diamond are deposited on oxidized silicon wafers via microwave plasma enhanced chemical vapour deposition (CVD), resulting in a diamond-on-insulator wafer, which can be used for fabricating waveguides and routing the light through photonic circuits [8,12]. The diamond layers are then polished via slurry based chemical-mechanical planarization (CMP) to a thickness of 600 nm. This leads to a surface roughness below 3 nm rms on a 25 $\mu m^2$ area, which is confirmed by atomic force microscopy (see AFM and SEM micrographs in Supplementary Fig. S1). This allows fabricating high quality integrated photonic circuits, which are pattered into a negative tone spin-on glass resist (HSQ 15 %) via electron beam lithography (JEOL 5300, 50kV). The patterns are transferred into diamond via dry etching by capacitively coupled reactive ion etching (RIE), using gas flows of 17 sccm Argon and 33 sccm Oxygen at a power of 200 W, leading to a diamond etch rate of 22 nm/min, with a selectivity of 2:1 for etching diamond versus etching the HSQ resist. After dry etching 300 nm into the diamond, the spin-on glass resist is removed using buffered oxide etch (BOE), resulting in pure diamond photonic structures ready for functionalization using DPN as illustrated in Fig.1a. This fabrication routine allows fabricating hundreds of devices per $cm^2$ of wafer material (see Supplementary Fig. S4 for a representative microscope image).

We fabricated waveguides, grating coupling structures and ring resonators to provide a rich template for later surface modification. By precisely aligning photonic components to the separation between neighboring DPN tips several devices can be addressed in parallel, allowing for surface-modifying each individual component with a different surfactant. Fabricated ring resonator devices coupled to a photonic bus waveguide are shown in Fig.1b. By choosing a particular mask layout arbitrary photonic devices can be inscribed into the diamond surface, while we here employ waveguides coupled to ring and microdisk resonators to realize both straight and curved optical elements. Scanning electron microscope (SEM) images reveal low surface roughness of the diamond waveguide top surface after nanofabrication. Small feature sizes down to 100 nm such as the gap between waveguide and microring resonators (Fig.1c) can be achieved with this fabrication method, while the limiting factor is only the thickness of the employed resist, which could be further reduced. Furthermore, the high etching anisotropy reached with our procedure results in near-vertical sidewalls. This is a result of the optimized RIE procedure carried out at low pressure and high RF power. We estimate from SEM micrographs that the sidewall angle is below 2° and therefore will enable even smaller gaps in principle. By connecting the so-fabricated waveguide based photonic circuits to on-chip grating structures contact-free coupling to optical fibers can be employed to realize a convenient readout platform for diamond integrated optical devices. This way a multitude of hybrid circuit devices can be fabricated and measured in a scalable fashion.

To surface modify pre-fabricated photonic devices, two different cantilever based approaches for functionalization were investigated. Firstly an AFM cantilever based deposition (DPN) which allows for drawing thin lines in arbitrary shapes (e.g. straight lines or circles). The second approach employs a microchannel cantilever based deposition, which allows depositing fine droplets at desired positions, for example in the inside of a microring resonator. Both approaches provide specific advantages, which will be discussed in later sections.

In the first case, L-DPN is used to functionalize photonic devices with a fluorescent lipid based ink. After carefully aligning the array of cantilevers to the waveguides via the machine's built in optical microscope (Fig.2a), the functionalization of several waveguides in parallel is performed by lowering the tips onto the waveguides and writing a 100 µm long lipid line. Fig.2b shows a fluorescence microscope image of one of the lines written in parallel. The overlay to a brightfield image of the device (Fig.2c), taken right after the fluorescence image shows the alignment of the photonic structure and the functionalization. AFM scans (Fig.2d) reveal the geometry of the waveguide and the functionalized area. The diamond waveguide has a width of 1 µm and a height of 300 nm and the line deposited by L-DPN can be clearly identified in detailed scan of a 300 x 300 nm$^2$ area on top of the waveguide. The cross section (Fig.2d Inset) shows that the deposited lines have a height of 4nm, corresponding to a dehydrated lipid bilayer, which can be written with a width down to 100 nm.

This parallelized functionalization can also be performed on arbitrarily shaped or curved photonic elements, which is shown here by functionalizing several rings connected to waveguides in a potential photonic circuit (Fig.3a). The circuit comprises four microring resonators, placed in the vicinity of a straight photonic waveguide on the top. The crosses visible above of the rings are used to perform the alignment of the cantilever array to the rings (see Supplementary Fig. S11 for an in-situ view). The microrings are fabricated with a pitch of 66 µm to match the distance between neighboring cantilevers during writing (see Supplementary Fig. S10 for details). Here, only three of the total twelve cantilever in such an array were coated with the fluorescent lipid ink. Thus when writing onto the microring structures, only the first three of the four rings in the structure were decorated with the fluorescent lipid ink in parallel, while the fourth remained as an unfunctionalized control device. The fluorescence measurement at the same position (Fig.3b) shows that three of the rings have been functionalized in parallel as intended, while the control device remains unchanged (see Supplementary Fig. S15 for further microscope images). The written rings are well aligned to the structured diamond substrate (Fig.3c) and the linewidth and height can be easily controlled by AFM scans as shown in Fig.3d. From the image it is apparent that the DPN written line is well guided along the radius of the rings on all three devices simultaneously, thus showcasing the high spatial precision available to the method. Here we only show the results for a single ring for clarity, while an overview of all three functionalized devices can be found in the supplementary information in Fig. S15.

A further convenient approach to functionalize nanophotonic structures is to apply droplet deposition by microchannel cantilevers. In this case we can profit from self-alignment occurring while the solvent of the ink formulation evaporates. A schematic of this is shown in Fig.4a: (I) Femtoliter droplets of the desired ink solutions are deposited into the ring resonator structure. (II) The solvent evaporates but is pinned at the outer rim of the ring structure by capillary forces. (III) Upon further evaporation of solvent, the droplet breaks up in the middle and turns into an ink ring structure. (IV) The ink ring opening increases, while the solvent evaporates and the solved compound is concentrated to the outer ring, finally leading to a well-defined deposition of the solved compound at the waveguide ring structure. In practice, the small volume droplet deposition into the ring structures was done with microchannel cantilevers [39]. By contacting the cantilever with the sample in the center of the ring structures, the dye solution was allowed to flow onto the sample and fills the diamond ring structure. Outside the DPN machine, in the laboratory humidity of about 30 % r.H., the solvent evaporates over the course of several minutes. While the ink solution dries out, the fluorescent dyes align themselves on the interior of the ring, as can be seen in series of fluorescence microscope images of the same device (Fig.4b). After 70 min, all solvent is evaporated leading to ring resonators with a perfectly aligned functionalized surface on the inside of the ring (Fig.4b IV). Furthermore, we performed this procedure with multiple different dyes in parallel: Three different fluorescent dyes are used in the experiment emitting in the red, green and blue wavelength range (see Supporting Information and Fig. S9 for the chemical structure). The three different ink solutions were deposited inside three adjacent rings by microchannel cantilevers as described above, while one empty ring acts as an unfunctionalized control. Fig.4c shows an overlay of a grey colour brightfield image of the diamond photonic circuit with three flourescence microscopy images in the different color channels. The resulting image shows that an accurate site specific functionalization of devices in close proximity is achieved by this combined top-down and self-assembly process.

In order to evaluate the performance of the photonic circuits for on-chip routing of fluorescence and its out-of-plane access for detection, we fabricated devices consisting of a waveguide of 1 µm width connected to a pair of in- and output focusing grating couplers [40], as shown in the inset of Fig.5a. We measured their transmission spectra using a broadband supercontinuum white light source covering ultraviolet up to near-infrared wavelengths and a spectrometer. By varying grating periods and fill factor of the Bragg grating, the coupling efficiency can be increased and the central coupling wavelength can be adjusted to the emission spectrum of a fluorescent dye. By changing the grating period from 250 nm to 380 nm at fill factors from 30-50% we can cover the full spectral range of visible light from 400 nm to 800 nm, as shown in Fig.5a. Each transmission peak corresponds to an individual measurement of a different device. The transmission decreases for smaller wavelengths, which can be mainly attributed to the increasing propagation losses at small wavelengths [41]. Due to the efficient polishing of our diamond thin films, the propagation losses are reduced compared to previous work (see Supplementary Fig. S6 and S7), showing the potential of the material used in this work. By employing these focusing grating couplers in photonic circuits light, coupled into the waveguides either from external sources or on-chip fluorescent emitters, can be guided to and extracted at arbitrary locations on chip.

In the second step the waveguide connecting the grating couplers is functionalized using DPN. After functionalizing the circuits, the emitted fluorescence is coupled into the waveguide, routed to a grating coupler and then scattered out-of-plane which allows it to be detected using a wide-field fluorescent microscope equipped with a sensitive camera. To show this, we functionalized the devices with a DPN line perpendicular across the waveguide and evaluated the amount of fluorescence at the grating couplers. Because the DPN written line is placed in the near-field of the photonic waveguide, light emitted by the die molecules will couple preferentially into the underlying diamond waveguide because of the higher refractive index compared to the surrounding air. Once emitted light is coupled into the diamond, it will be propagating along the waveguide towards the grating coupler and then emitted out of plane through Bragg scattering. To distinguish between guided light and residual false counts on the camera, the fluorescent background due to stray light, detector dark counts and residual excitation light passing the optical filters is removed from the measured signal. We first take reference measurements on unfunctionalized diamond samples. In the red channel of the imaging microscope the bare diamond reference substrate shows a 30% higher fluorescence signal than obtained with oxidized silicon samples. We attribute this background fluorescence to impurities in diamond, such as nitrogen vacancy color centers which are commonly encountered in CVD grown diamond layers, especially at the grain boundaries. Therefore also unfunctionalized waveguides show fluorescence which is coupled out at the grating couplers (Fig.5b). In case of a functionalized device the light emitted by the red fluorescent dye is routed and scattered out-of-plane at the grating couplers in the same fashion, which makes it challenging to distinguish the contributions from the DPN functionalization and the background. As the diamond background fluorescence is lower in the blue spectral range, we use the blue fluorescent dye (Cascade Blue) to quantify the fluorescence coupled out at the grating couplers. As shown in Fig.5c, the light gets primarily scattered at the first few grating lines, similar to light from integrated visible light sources on silicon nitride substrates [42]. We furthermore functionalize aluminum nitride nanophotonic circuits via the same DPN protocol, in order to have a different waveguide material as a reference (see Supplementary Fig. S12 and S13). By pairwise measurement and comparison of the couplers of a functionalized and unfunctionalized diamond device, as well as four couplers of unfunctionalized control devices far away from the functionalized devices, the amount of coupled fluorescence light from the functionalization can be established. We integrate the detected light across an area (indicated with the yellow line) on the grating coupler, to account for stray light. The local background was estimated by integrating over the same area directly below the respective coupler to allow for a background subtraction (See Supplementary Fig. S14 for details). For the functionalized devices we measure intensity (a.u.) of $1025 \pm 43$ in comparison to $760 \pm 13$ for the direct neighboring couplers on unfunctionalized devices and $792 \pm 169$ for the distant control couplers (See Supplementary Table S1 for all measurement values). Thus, couplers on the functionalized device emit about 35% higher intensity than on the neighboring unfunctionalized devices, showing that the fluorescence from the functionalization of the diamond waveguide can be routed inside photonic circuits to focusing grating couplers. This allows detecting the fluorescence signal at a place of choice on the chip, enabling out-of-plane access for excitation and detection.

The nanofunctionalization approach presented here allows for labeling arbitrary photonic components with high precision and accuracy. Employing DPN writing in conjecture with planar nanofabrication techniques thus enables site-specific coloring of multiple nanophotonic systems in parallel, without immersing the entire wafer in a liquid or labelactive environment. This is of particular importance when fragile devices need to be surface modified or in the case of several functionalization steps required in succession. Using DPN, multiple writing sessions can be performed with varying dies, thus allowing for multiplexed functionalization and also the creation of complex patterns on a sample surface. By combining nanophotonic circuits with precision fluorescent labelling optical signal processing of light emitted from precisely defined locations on chip can thus be achieved. The use of diamond as target material in this context paves the road towards biocompatible optical devices that can be operated in a wide wavelength range and also in a wide range of robust biomedical and environmental high throuhgput sensing applications. In the nanophotonic circuits we show that waveguiding from near-infrared wavelengths down to the ultraviolet wavelength regime can be achieved on chip, thus covering important spectral regions for fluorescent imaging. Exploiting the full flexibility of scalable nanofabrication routines provided by both ebeam lithography and DPN, our approach allows for realizing a multitude of functional devices in a controllable fashion, such as extending the functionalization to different chemistries for sensing purposes or realizing nanoscale light emitters on chip, thus paving the way towards high-throughput chipscale solutions for biophotonic sensing.

**Experimental Section**

*Plasma enhanced CVD deposition of diamond:*

To initiate the diamond growth, a diamond nano-particle seed layer is deposited onto the substrate by ultrasonification for 30 minutes in a water based suspension of ultra-dispersed (0.1 wt %) nano-diamond particles of typically 5-10 nm size [43]. The samples are then rinsed with deionized water and methanol. After dry blowing, the wafer is transferred into an ellipsoidal microwave plasma reactor [44]. The diamond films are then grown using 1 % CH4 in 99 % H2, at a pressure of 55 mbar, a microwave power of 3.5 kW and a temperature of 850 °C. In order to avoid angular non-uniformities arising from the gas flow the substrate is continuously rotated. Growth rates are in the range of 1-2 µm/h. After growth, the samples are cleaned in concentrated $HNO_3:H_2SO_4$ to remove surface contaminants.

*Chemical-mechanical planarization (CMP):* For slurry based CMP of diamond thin films [45] a contact force of 120 N is applied and about 80 ml/min polishing liquid is used at a rotational frequency of 90 1/min. The as-deposited diamond film of 900 nm thickness shows a roughness of 35nm rms. After removing material and reducing the film thickness by CMP to 600 nm, the surface roughness below 3 nm rms on a 25 $\mu m^2$ area, which is confirmed by atomic force microscopy.

*Site-specific functionalization:* The site-specific functionalization of the devices by DPN and microchannel cantilevers was performed on a DPN5000 system (NanoInk) and a NLP2000 system (NanoInk), respectively. For DPN, cantilever arrays of M-type were used (Advanced Creative Solutions Technology (ACST)) with a pitch of either 66 µm or 100 µm between adjacent cantilevers were employed. Phospholipid based formulations were used as functional inks: 1,2-dioleoyl-sn-glycero-3-phosphocholine (DOPC) was used as carrier and admixed with 1 mol% of 1,2-dioleoyl-sn-glycero-3-phosphoethanolamine-N-(lissamine rhodamine B sulfonyl) (ammonium salt) (Rho-PE) to render the mixture fluorescent with red emission (both compounds from Avanti Polar Lipids). For the chemical structure of the phospholipids see Fig. S8 in the Supporting Information. To cover the cantilever arrays with the phospholipid mixture, Inkwells (ACST) were loaded with 1-2 µl of the lipid mixture in each reservoir at a concentration of 20 mg/ml in chloroform. After evaporation of the chloroform in a desiccator for 15 min, the cantilever arrays were dipped into the inkwells at a humidity of 70 % r.H. for 5 min. Excess ink on the cantilevers was removed by manually drawing line patterns on a sacrificial area of the sample. After this, the cantilevers were carefully aligned with the device structures and the desired line pattern was written. Writing took place at controlled relative humidity ranging from 30 to 40 % r.H. and the writing speed was 5 µm/s to 10 µm/s depending on desired line thickness.

For spotting by microchannel cantilevers, cantilevers (SPT-S-C10S, Bioforce Nanosciences) were plasma cleaned by 10 sccm oxygen plasma at 100 mTorr, 30 W for 2 min, then attached to the DPN system's tip holder by double sided sticky tape and filled with 1 µl of the desired ink solution. Azide modifications of TAMRA, Alexa 488, and Cascade Blue (all from Life Technologies) were dissolved in DI water at concentrations of 1 mg/ml and admixed with glycerol (87% in DI water) in 7 parts dye solution to 3 parts glycerol to prevent premature drying of the ink in the microchannel cantilevers reservoir. Spotting was performed at 60 % r.H. and with an additional inclination of the DPN system's sample stage by 8° to prevent contact of the reservoir with the substrate.

*Fluorescence microscopy:* For quantifying the fluorescence a wide-field fluorescent microscope (Eclipse 80i upright fluorescence microscope) is used, equipped with a sensitive camera (CoolSNAP HQ 2 camera (Photometrics)). The broadband excitation light source (Intensilight illumination (Nikon)) is combined with standard sets of filters (TexasRed (red), FITC (green), DAPI (blue), Nikon) to separate excitation and emission spectra, depending on the used dye molecule.

*Atomic force microscopy:* Atomic force microscopy was performed on a Dimension Icon AFM (Bruker) in standard tapping mode in air. Cantilevers used where of type NSC15 (MikroMasch) with a nominal force constant of 46 N/m and a resonance frequency of 325 kHz.

*Optical transmission measurements:*
The transmission through on-chip nanophotonic circuits is measured using a custom measurement setup for precision alignment and multi-port optical access (see Supplementary Fig. S5 for a schematic of the setup). Light from an unpolarized supercontinuum source (Leukos SM-30-UV) is sent through the on-chip devices via focusing grating couplers and detected on a second coupling port with a fiber coupled spectrometer (Ocean Optics JAZ). Both instruments are connected to an optical fiber-array. The fiber-array consists of a matrix of optical fibers, with a fixed spacing of 250 µm between individual fibers. The distance between the focusing grating couplers of the fabricated devices is therefore 250 µm as well. The chip is mounted on an computer controlled x/y/z- piezo-stage, which allows to easily align the fiber-to-chip coupling for many devices in a short time with precision better than 30nm.

**Acknowledgements**

We acknowledge support by DFG grant PE 1832/1-1 and PE 1832/2-1. P. R. acknowledges support by the Karlsruhe School of Optics and Photonics (KSOP) and the Deutsche Telekom Stiftung. We also appreciate support by the Deutsche Forschungsgemeinschaft (DFG) and the State of Baden-Württemberg through the DFG-Center for Functional Nanostructures (CFN) within subproject A6.4. This work was partly carried out with the support of the Karlsruhe Nano Micro Facility (KNMF, http://www.knmf.kit.edu), a Helmholtz Research Infrastructure at Karlsruhe Institute of Technology (KIT, http://www.kit.edu). The authors further wish to thank Silvia Diewald and Stefan Kühn for assistance in device fabrication.

**References**

[1] A. Faraon, C. Santori, Z. Huang, V. Acosta, & R. Beausoleil, *Phys. Rev. Lett.*, **2012**, *109*, 033604

[2] B. J. M. Hausmann, B. Shields, Q. Quan, P. Maletinsky, M. McCutcheon, J. T. Choy, T. M. Babinec, A. Kubanek, A. Yacoby, M. D. Lukin, M. Loncar, *Nano letters* **2012**, *12*, 1578.

[3] J. Riedrich-Moeller, L. Kipfstuhl, C. Hepp, E. Neu, C. Pauly, F. Mücklich, A. Baur, M. Wandt, S. Wolff, M. Fischer, S. Gsell, M. Schreck, C. Becher, *Nature Nanotechnology* **2012**, *7*, 69.

[4] B. J. M. Hausmann, B. J. Shields, Q. Quan, Y. Chu, N. P. de Leon, R. Evans, M. J. Burek, A. S. Zibrov, M. Markham, D. J. Twitchen, H. Park, M. D. Lukin, M. Lonc R, *Nano letters* **2013**, *13*, 5791.

[5] N. Vermeulen, J. E. Sipe, L. G. Helt, H. Thienpont, *Laser & Photonics Reviews* **2012**, *6*, 793.

[6] B. J. M. Hausmann, I. Bulu, V. Venkataraman, P. Deotare, M. Lončar, *Nature Photonics* **2014**, *8*, 369.

[7] P. Maletinsky, S. Hong, M. S. Grinolds, B. Hausmann, M. D. Lukin, R. L. Walsworth, M. Loncar, A. Yacoby, *Nature nanotechnology* **2012**, *7*, 320.

[8] P. Rath, S. Khasminskaya, C. Nebel, C. Wild, W. H. P. Pernice, *Nature Communications* **2013**, *4*, 1690.

[9] B. Zhang, Y. Li, C.-Y. Fang, C.-C. Chang, C.-S. Chen, Y.-Y. Chen, H.-C. Chang, *Small* **2009**, *5*, 2716.

[10] N. Mohan, C.-S. Chen, H.-H. Hsieh, Y.-C. Wu, H.-C. Chang, *Nano letters* **2010**, *10*, 3692.

[11] T. Meinhardt, D. Lang, H. Dill, A. Krueger, *Advanced Functional Materials* **2011**, *21*, 494.

[12] P. Rath, N. Gruhler, S. Khasminskaya, C. Nebel, C. Wild, W. H. P. Pernice, *Optics Express* **2013**, *21*, 11031.

[13] Y. Chen, W. S. Fegadolli, W. M. Jones, A. Scherer, M. Li, *ACS nano* **2013**, *8*, 522–527.

[14] M. Iqbal, M. A. Gleeson, B. Spaugh, F. Tybor, W. G. Gunn, M. Hochberg, T. Baehr-Jones, R. C. Bailey, L. C. Gunn, *IEEE Journal of Selected Topics in Quantum Electronics* **2010**, *16*, 654.

[15] J. T. Robinson, L. Chen, M. Lipson, *Optics Express* **2008**, *16*, 4296.

[16] C. Subramani, Y. Ofir, D. Patra, B. J. Jordan, I. W. Moran, M.-H. Park, K. R. Carter, V. M. Rotello, *Adv. Funct. Mater.* **2009**, *19*, 2937.

[17] C. Subramani, N. Cengiz, K. Saha, T. N. Gevrek, X. Yu, Y. Jeong, A. Bajaj, A. Sanyal, V. M. Rotello, *Adv. Mater.* **2011**, *23*, 3165.

[18] Y. Ofir, I. W. Moran, C. Subramani, K. R. Carter, V. M. Rotello, *Adv. Mater.* **2010**, *22*, 3608.

[19] R. D Piner, J. Zhu, F. Xu, S. Hong, C. A. Mirkin, *Science* **1999**, *283*, 661.

[20] C.-C. Wu, D. N. Reinhoudt, C. Otto, V. Subramaniam, A. H. Velders, *Small* **2011**, *7*, 989.

[21] H. Zhang, S.-W. Chung, C. A. Mirkin, *Nano Letters* **2003**, *3*, 43.

[22] D. J. Pena, M. P. Raphael, J. M. Byers, *Langmuir* **2003**, *19*, 9028.

[23] A. Noy, A. E. Miller, J. E. Klare, B. L. Weeks, B. W. Woods, J. J. DeYoreo, *Nano Letters* **2002**, *2*, 109.

[24] J.-H. Lim, C. A. Mirkin, *Advanced Materials* **2002**, *14*, 1474.

[25] L. M. Demers, D. S. Ginger, S.-J. Park, Z. Li, S.-W. Chung, C. A. Mirkin, *Science* **2002**, *296*, 1836.

[26] G. Agarwal, R. R. Naik, M. O. Stone, *Journal of the American Chemical Society* **2003**, *125*, 7408.

[27] M. Ben Ali, T. Ondarçuhu, M. Brust, C. Joachim, *Langmuir* **2002**, *18*, 872.

[28] J. C. Garno, Y. Yang, N. A. Amro, S. Cruchon-Dupeyrat, S. Chen, G.-Y. Liu, *Nano Letters* **2003**, *3*, 389.

[29] S. Lenhert, P. Sun, Y. Wang, H. Fuchs, C. A. Mirkin, *Small* **2007**, *3*, 71.

[30] M. Hirtz, R. Corso, S. Sekula-Neuner, H. Fuchs, *Langmuir* **2011**, *27*, 11605.

[31] S. Sekula-Neuner, J. Maier, E. Oppong, A. C. B. Cato, M. Hirtz, H. Fuchs, *Small* **2012**, *8*, 585.

[32] E. Oppong, P. N. Hedde, S. Sekula-Neuner, L. Yang, F. Brinkmann, R. M. Dörlich, M. Hirtz, H. Fuchs, G. U. Nienhaus, A. C. B. Cato, *Small* **2014**, *10*, 1991.

[33] M. Hirtz, A. Oikonomou, T. Georgiou, H. Fuchs, A. Vijayaraghavan, *Nature communications* **2013**, *4*, 2591.

[34] U. Bog, T. Laue, T. Grossmann, T. Beck, T. Wienhold, B. Richter, M. Hirtz, H. Fuchs, H. Kalt, T. Mappes, *Lab on a chip* **2013**, *13*, 2701.

[35] M. Hirtz, M. Lyon, W. Feng, A. E. Holmes, H. Fuchs, P. A. Levkin, *Beilstein journal of nanotechnology* **2013**, *4*, 377.

[36] M. Hirtz, A. M. Greiner, T. Landmann, M. Bastmeyer, H. Fuchs, *Advanced Materials Interfaces* **2014**, DOI 10.1002/admi.201300129.

[37] A. Krueger, *Chemistry* **2008**, *14*, 1382.

[38] V. N. Mochalin, O. Shenderova, D. Ho, Y. Gogotsi, *Nature nanotechnology* **2012**, *7*, 11.

[39] J. Xu, M. Lynch, J. L. Huff, C. Mosher, S. Vengasandra, G. Ding, E. Henderson, *Biomedical microdevices* **2004**, *6*, 117.

[40] D. Taillaert, W. Bogaerts, P. Bienstman, T. F. Krauss, P. Van Daele, I. Moerman, S. Verstuyft, K. De Mesel, R. Baets, *IEEE Journal of Quantum Electronics* **2002**, *38*, 949.

[41] M. Stegmaier, J. Ebert, J. M. Meckbach, K. Ilin, M. Siegel, W. H. P. Pernice, *Applied Physics Letters* **2014**, *104*, 091108.

[42] S. Khasminskaya, F. Pyatkov, B. S. Flavel, W. H. Pernice, R. Krupke, Advanced Materials **2014**, *26*, 3465–3472.

[43] O. A. Williams, O. Douhéret, M. Daenen, K. Haenen, E. Ōsawa, M. Takahashi, *Chemical Physics Letters* **2007**, *445*, 255.

[44] M. Fuener, C. Wild, P. Koidl, *Applied Physics Letters* **1998**, *72*, 1149.

[45] E. L. H. Thomas, G. W. Nelson, S. Mandal, J. S. Foord, O. A. Williams, *Carbon* **2014**, *68*, 473.

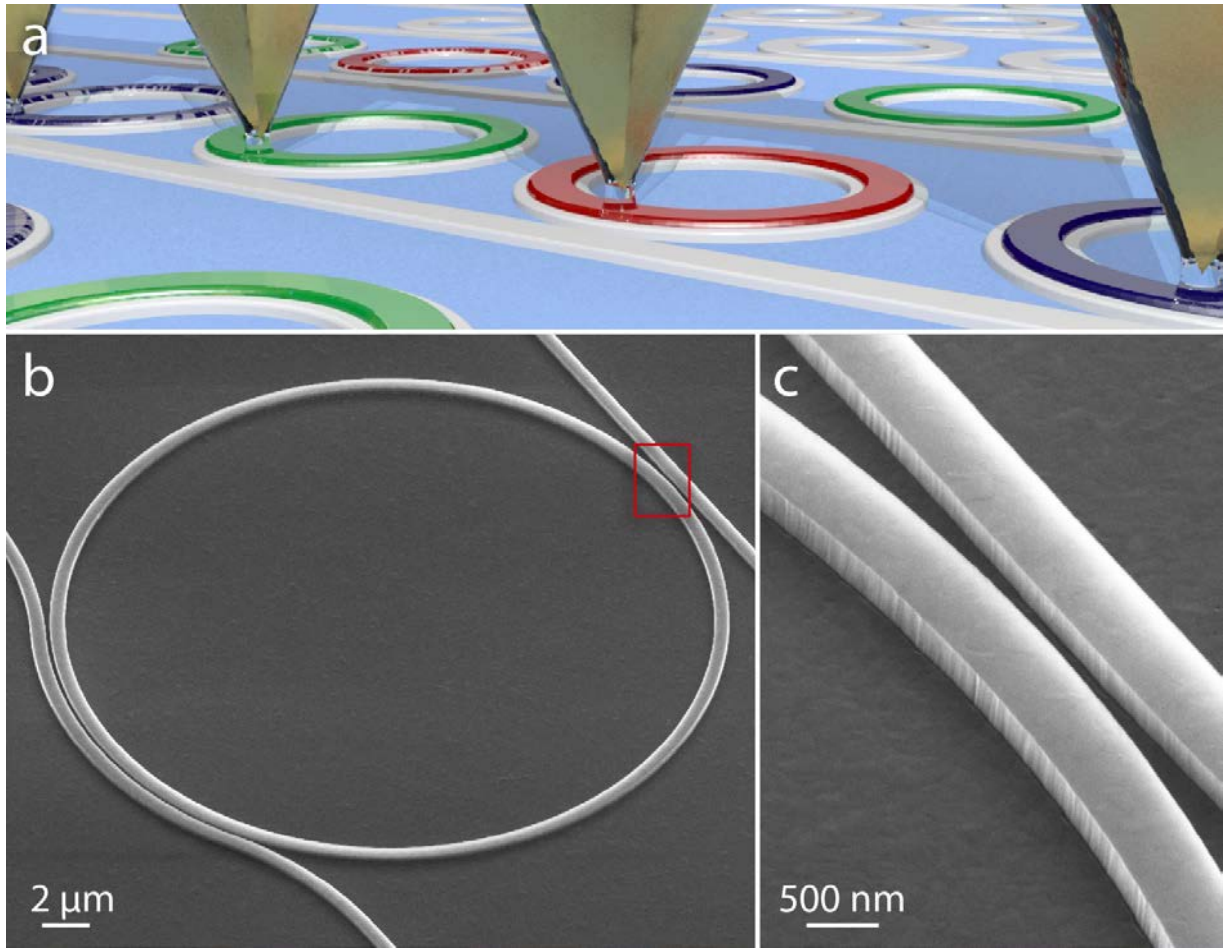

**Figure 1: Surface functionalization of diamond nanophotonic components.**

(a) Illustration of the surface functionalization of diamond photonic components by dip-pen nanolithography. Multiple devices get simultaneously functionalized employing AFM-cantilever arrays. (b) SEM micrograph of fabricated diamond photonic waveguides coupled to a ring resonator. (c) SEM image showing the smooth top surface due to chemical-mechanical planarization and the straight sidewalls due to reactive ion etching.

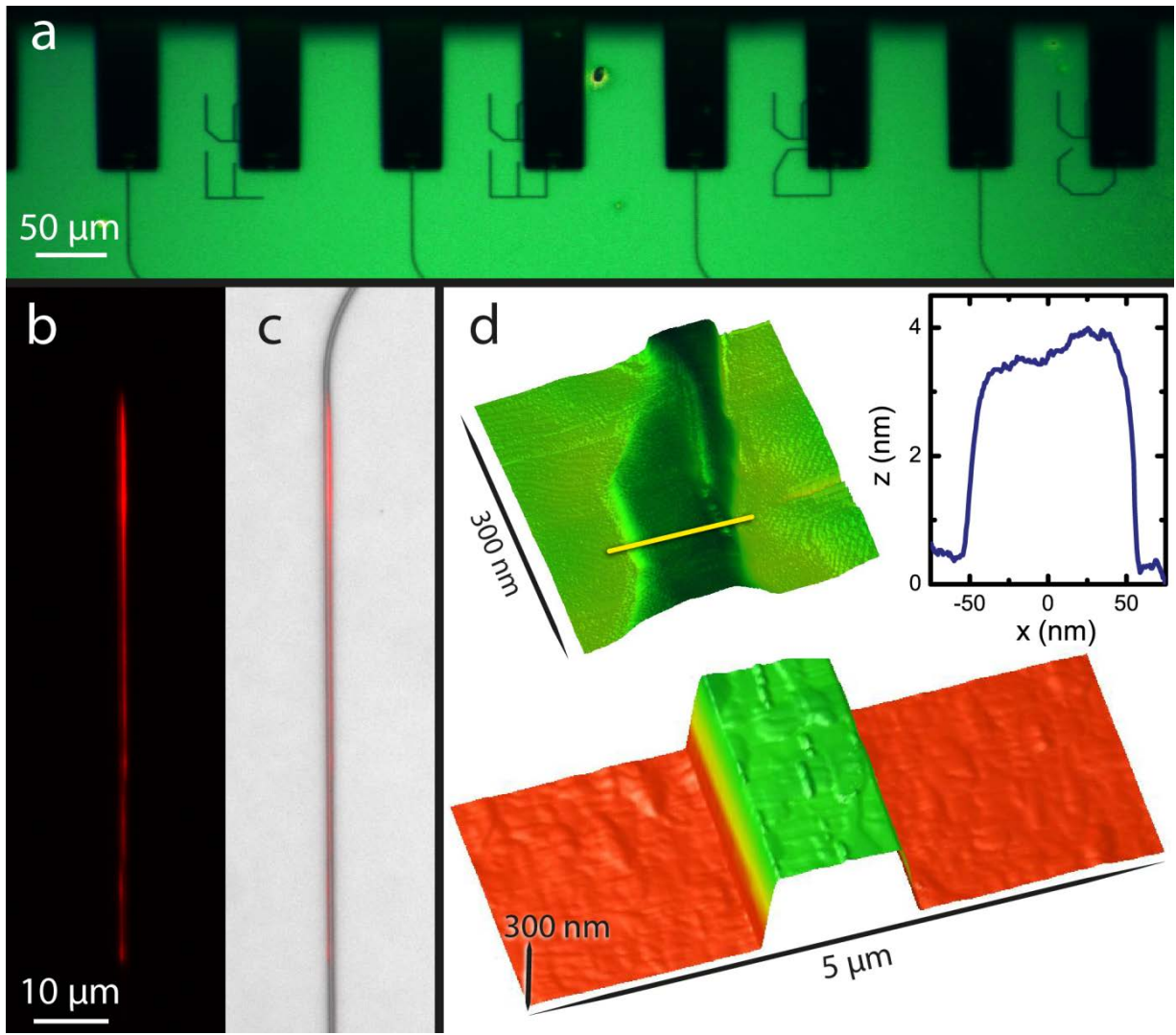

**Figure 2: Parallel DPN functionalization and alignment.**

(a) Microscope picture showing the parallel alignment of the AFM cantilevers for DPN to the waveguides. (b) False colour image of the measured fluorescence. (c) Fluorescence image as an overlay of a brightfield microscope picture showing the alignment result of the DPN process. (d) AFM micrograph of the diamond waveguide. The insets show an AFM micrograph and the height profile of the lipid ink line on the waveguide.

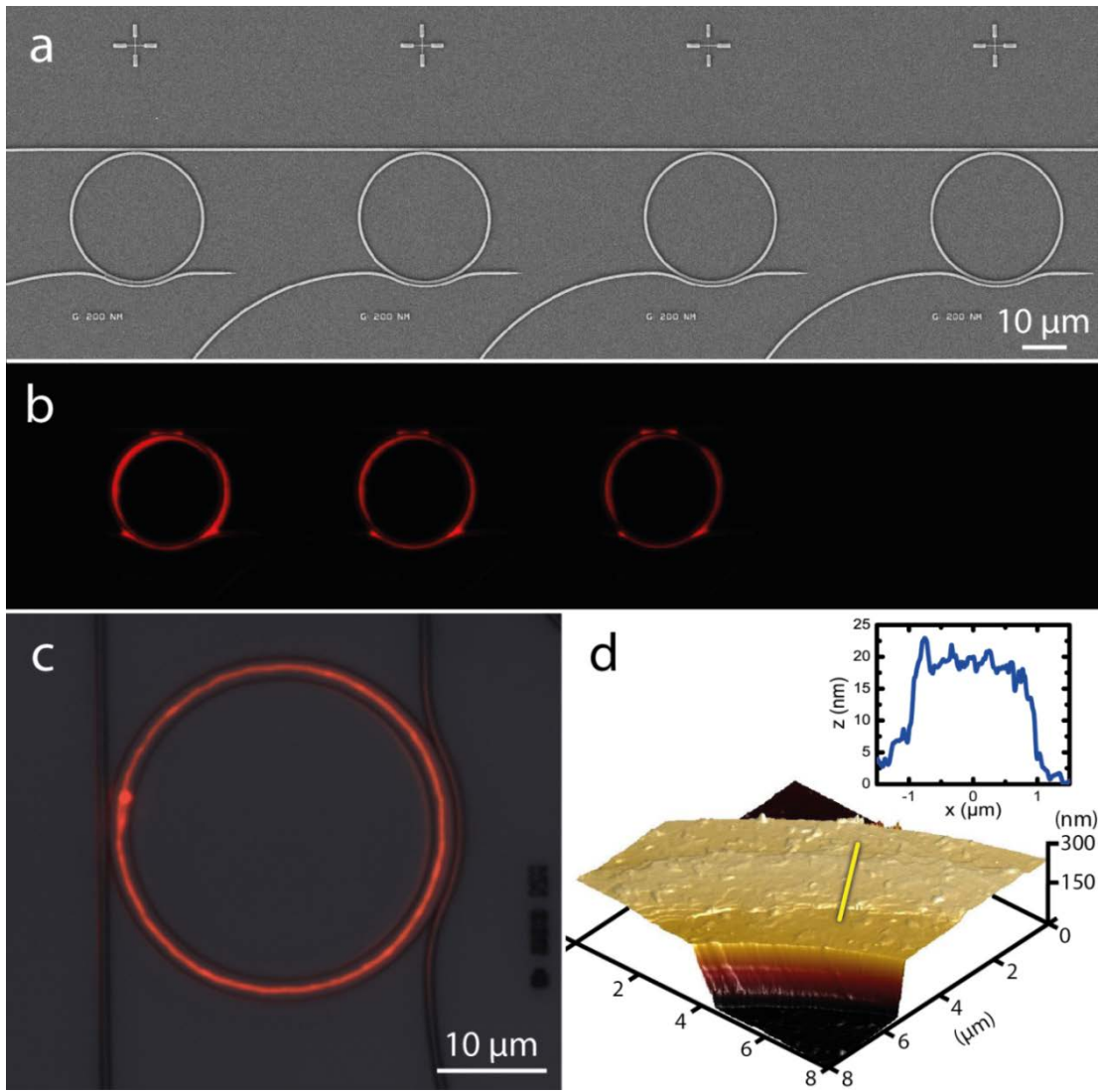

**Figure 3: DPN functionalization of curved photonic elements.**
(a) SEM micrograph showing the geometry of several rings connected to waveguides in a potential photonic circuit. (b) Flourescence measurement of three rings that were functionalized in parallel and one empty ring as a control device, showing the potential to functionalize curved structures. (c) Flourescence measurement on one ring as an overlay to the brightfield image showing the good alignment of DPN to the structured substrate (d) AFM micrograph showing the lipid line written by L-DPN. The inset shows a cross section revealing 2 µm width and about 20 nm height of this particular line.

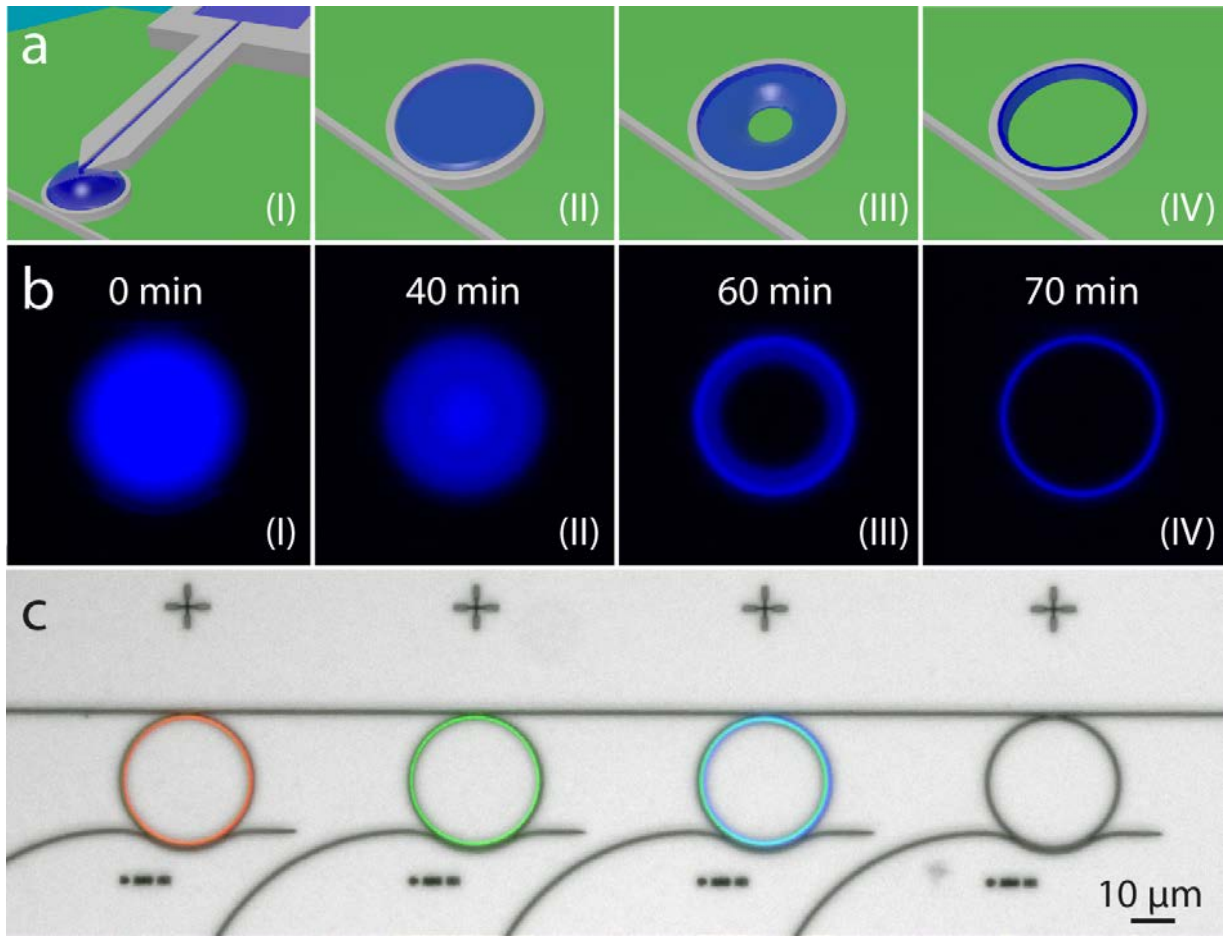

**Figure 4: Self-aligned functionalization of diamond ring resonators.**

(a) Schematic of droplet deposition into the ring structures (b) Series of false colour fluorescence images of the ring functionalized with Cascade Blue ink showing the drying process which leads to a self-alignment on the inside of the ring (c) Overlay of the three fluorescence channels on a brightfield microscope micrograph showing the result of three rings functionalized at the same time with three different fluorescent inks (from left to right: TAMRA, Alexa488, Cascade Blue). The right ring is a control, showing no fluorescence.

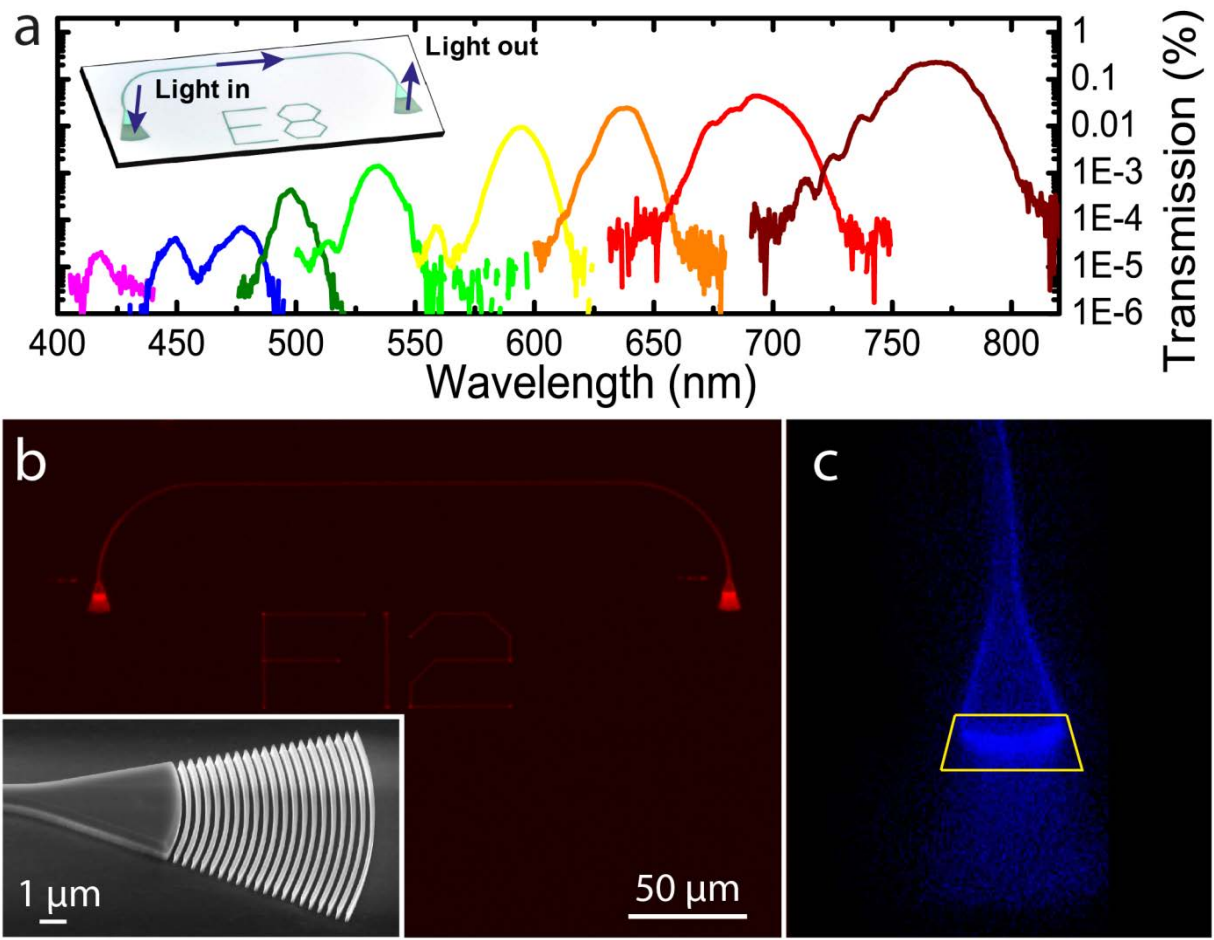

**Figure 5: Routing of fluorescent emission with on-chip waveguides.**

(a) Transmission spectra of several diamond waveguides coupled to in- and output grating couplers. The central coupling wavelength of the transmission spectrum can be tuned through the visible spectral range. Inset: Sketch showing the device geometry and the principle of the transmission measurement (b) Fluorescence microscopy image of an unfunctionalized waveguide, showing the autofluorescence in the red spectral range. Inset: SEM micrograph of one diamond grating coupler. (c) Fluorescence microscopy image of a grating coupler of a device functionalized with Cascade Blue showing that the guided light is scattered out-of-plane mainly by the first grating lines.